\documentclass[]{spie}  
\usepackage{ifthen}
\usepackage[]{graphicx}
\usepackage{subfigure}
\usepackage{mdwlist}
\usepackage{floatrow}
\DeclareGraphicsExtensions{.pdf,.png,.jpg}

\def\lcdm{$\Lambda$CDM }
\def\fmux{fMux }
\def\psat{$P_{\mathrm{sat}}$ }
\title{Year two instrument status of the SPT-3G cosmic microwave
background receiver}

\author[a,b]{A.~N.~Bender}
\author[c]{P.~A.~R.~Ade}
\author[d,e]{Z.~Ahmed}
\author[f,b]{A.~J.~Anderson}
\author[g]{J.~S.~Avva}
\author[h]{K.~Aylor}
\author[b]{P.~S.~Barry}
\author[b]{R.~Basu Thakur}
\author[f,b,i]{B.~A.~Benson}
\author[a,b]{L.~S.~Bleem}
\author[j,a]{S.~Bocquet}
\author[a]{K.~Byrum}
\author[b,k,l,a,i]{J.~E.~Carlstrom}
\author[a,b]{F.~W.~Carter}
\author[a]{T.~W.~Cecil}
\author[a,b,i]{C.~L.~Chang}
\author[e]{H.-M.~Cho}
\author[m]{J.~F.~Cliche}
\author[b,i]{T.~M.~Crawford}
\author[g]{A.~Cukierman}
\author[g]{T.~de~Haan}
\author[n]{E.~V.~Denison}
\author[o]{J.~Ding}
\author[m,p]{M.~A.~Dobbs}
\author[q]{S.~Dodelson}
\author[b,l]{D.~Dutcher}
\author[r]{W.~Everett}
\author[s]{A.~Foster}
\author[b,t]{J.~Gallicchio}
\author[m]{A.~Gilbert}
\author[g]{J.~C.~Groh}
\author[g]{S.~T.~Guns}
\author[r,u]{N.~W.~Halverson}
\author[v,a]{A.~H.~Harke-Hosemann}
\author[g]{N.~L.~Harrington}
\author[a,b]{J.~W.~Henning}
\author[n]{G.~C.~Hilton}
\author[m,p,w]{G.~ P.~Holder}
\author[g]{W.~L.~Holzapfel}
\author[g]{N.~Huang}
\author[d,x,e]{K.~D.~Irwin}
\author[g]{O.~B.~Jeong}
\author[f]{M.~Jonas}
\author[o]{T.~S.~Khaire}
\author[h]{L.~Knox}
\author[v]{A.~M.~Kofman}
\author[s]{M.~Korman}
\author[f]{D.~L.~Kubik}
\author[a]{S.~Kuhlmann}
\author[d,x,e]{C.-L.~Kuo}
\author[g,y]{A.~T.~Lee}
\author[b,i]{E.~M.~Leitch}
\author[b]{A.~E.~Lowitz}
\author[b,k,l,i]{S.~S.~Meyer}
\author[z]{D.~Michalik}
\author[m]{J.~Montgomery}
\author[v]{A.~Nadolski}
\author[aa]{T.~Natoli}
\author[f]{H.~Ngyuen}
\author[m]{G.~I.~Noble}
\author[o]{V.~Novosad}
\author[b]{S.~Padin}
\author[b,l]{Z.~Pan}
\author[o]{J.~Pearson}
\author[o]{C.~M.~Posada}
\author[b,l]{W.~Quan}
\author[bb]{S.~Raghunathan}
\author[f,b]{A.~Rahlin}
\author[bb]{C.~L.~Reichardt}
\author[s]{J.~E.~Ruhl}
\author[r]{J.T.~Sayre}
\author[b,i]{E.~Shirokoff}
\author[cc]{G.~Smecher}
\author[b,l]{J.~A.~Sobrin}
\author[dd]{A.~A.~Stark}
\author[d,x]{K.~T.~Story}
\author[y]{A.~Suzuki}
\author[d,x,e]{K.~L.~Thompson}
\author[c]{C.~Tucker}
\author[n]{L.~R.~Vale}
\author[aa,ee]{K.~Vanderlinde}
\author[v,w]{J.~D.~Vieira}
\author[a]{G.~Wang}
\author[ff,g]{N.~Whitehorn}
\author[b]{W.~L.~K.~Wu}
\author[a]{V.~Yefremenko}
\author[d,x,e]{K.~W.~Yoon}
\author[ee]{M.~R.~Young}

\affil[a]{High-Energy Physics Division, Argonne National Laboratory, 9700 South Cass Avenue., Argonne, IL, USA 60439}
\affil[b]{Kavli Institute for Cosmological Physics, University of Chicago, 5640 South Ellis Avenue, Chicago, IL, USA 60637}
\affil[c]{School of Physics and Astronomy, Cardiff University, Cardiff CF24 3YB, United Kingdom}
\affil[d]{Kavli Institute for Particle Astrophysics and Cosmology, Stanford University, 452 Lomita Mall, Stanford, CA, USA 94305}
\affil[e]{SLAC National Accelerator Laboratory, 2575 Sand Hill Road, Menlo Park, CA, USA 94025}
\affil[f]{Fermi National Accelerator Laboratory, MS209, P.O. Box 500, Batavia, IL, USA 60510}
\affil[g]{Department of Physics, University of California, Berkeley, CA, USA 94720}
\affil[h]{Deptartment of Physics, University of California, One Shields Avenue, Davis, CA 95616}
\affil[i]{Department of Astronomy and Astrophysics, University of Chicago, 5640 South Ellis Avenue, Chicago, IL, USA 60637}
\affil[j]{Ludwig-Maximilians-Universit{\"a}t, Scheiner- str. 1, 81679 Munich, Germany}
\affil[k]{Enrico Fermi Institute, University of Chicago, 5640 South Ellis Avenue, Chicago, IL, USA 60637}
\affil[l]{Department of Physics, University of Chicago, 5640 South Ellis Avenue, Chicago, IL, USA 60637}
\affil[m]{Department of Physics, McGill University, 3600 Rue University, Montreal, Quebec H3A 2T8, Canada}
\affil[n]{NIST Quantum Devices Group, 325 Broadway Mailcode 817.03, Boulder, CO, USA 80305}
\affil[o]{Materials Sciences Division, Argonne National Laboratory, 9700 South Cass Avenue, Argonne, IL, USA 60439}
\affil[p]{Canadian Institute for Advanced Research, CIFAR Program in Cosmology and Gravity, Toronto, ON, M5G 1Z8, Canada}
\affil[q]{Department of Physics, Carnegie Mellon University, Pittsburgh, Pennsylvania, USA 15312}
\affil[r]{CASA, Department of Astrophysical and Planetary Sciences, University of Colorado, Boulder, CO, USA 80309}
\affil[s]{Department of Physics, Center for Education and Research in Cosmology and Astrophysics, Case Western Reserve University, Cleveland, OH, USA 44106}
\affil[t]{Harvey Mudd College, 301 Platt Boulevard., Claremont, CA, USA 91711}
\affil[u]{Department of Physics, University of Colorado, Boulder, CO, USA 80309}
\affil[v]{Department of Astronomy, University of Illinois at Urbana-Champaign, 1002 West Green Street, Urbana, IL, USA 61801}
\affil[w]{Department of Physics, University of Illinois Urbana-Champaign, 1110 West Green Street, Urbana, IL, USA 61801}
\affil[x]{Deptartment of Physics, Stanford University, 382 Via Pueblo Mall, Stanford, CA, USA 94305}
\affil[y]{Physics Division, Lawrence Berkeley National Laboratory, Berkeley, CA, USA 94720}
\affil[z]{University of Chicago, 5640 South Ellis Avenue, Chicago, IL, USA 60637}
\affil[aa]{Dunlap Institute for Astronomy \& Astrophysics, University of Toronto, 50 St. George Street, Toronto, ON, M5S 3H4, Canada}
\affil[bb]{School of Physics, University of Melbourne, Parkville, VIC 3010, Australia}
\affil[cc]{Three-Speed Logic, Inc., Vancouver, B.C., V6A 2J8, Canada}
\affil[dd]{Harvard-Smithsonian Center for Astrophysics, 60 Garden Street, Cambridge, MA, USA 02138}
\affil[ee]{Department of Astronomy \& Astrophysics, University of Toronto, 50 St. George Street, Toronto, ON, M5S 3H4, Canada}
\affil[ff]{Department of Physics and Astronomy, University of California, Los Angeles, CA, USA 90095}

\authorinfo{Send correspondence to A. N. Bender: abender@anl.gov}

\begin{document}
\maketitle

\begin{abstract}
The South Pole Telescope (SPT) is a millimeter-wavelength telescope
designed for high-precision measurements of the cosmic microwave background (CMB).   The SPT measures
both the temperature and polarization of the CMB with a large
aperture, resulting in high resolution maps sensitive to signals
across a wide range of angular scales on the sky.  With these
data, the SPT has the potential to make a broad range of cosmological
measurements.  These include constraining the effect of
massive neutrinos on large-scale structure formation as well as cleaning galactic and cosmological foregrounds from CMB polarization data in future searches for inflationary gravitational waves.
The SPT began observing in January  2017 with a new receiver (SPT-3G) containing $\sim$16,000
polarization-sensitive transition-edge sensor bolometers. Several key
technology developments have enabled this large-format focal plane,
including advances in detectors, readout electronics, and large
millimeter-wavelength optics.  We discuss the implementation of these
technologies in the SPT-3G receiver as well as the challenges they
presented.  In late 2017 the implementations of all three of these technologies were modified to optimize total performance.   Here, we present the current instrument status of the SPT-3G receiver.
\end{abstract}

\section{Introduction}
\begin{figure}[t]\centering
\subfigure{\includegraphics[height=0.25\textheight]{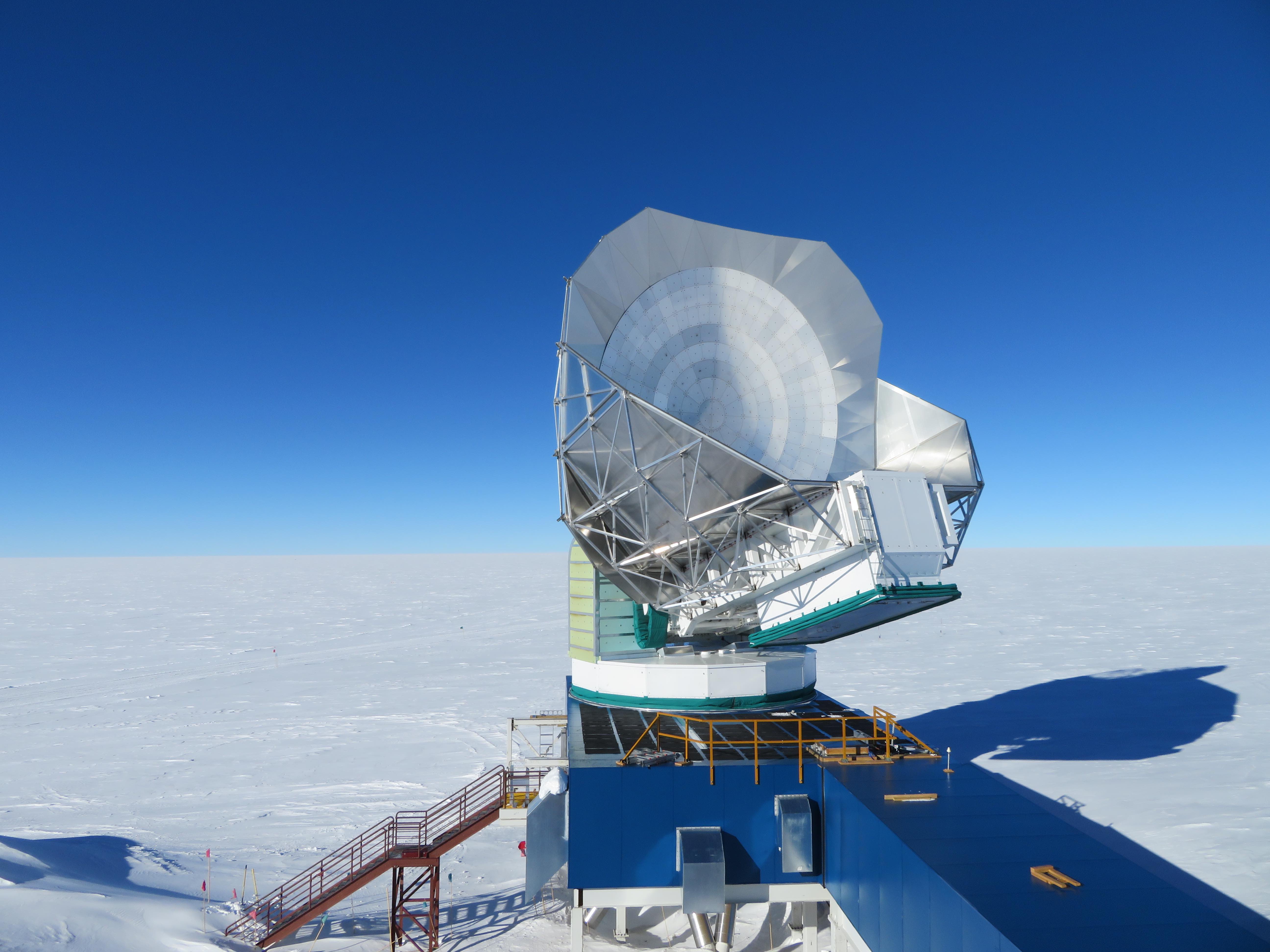}}
\hspace{0.03\textwidth}
\subfigure{\includegraphics[height=0.25\textheight]{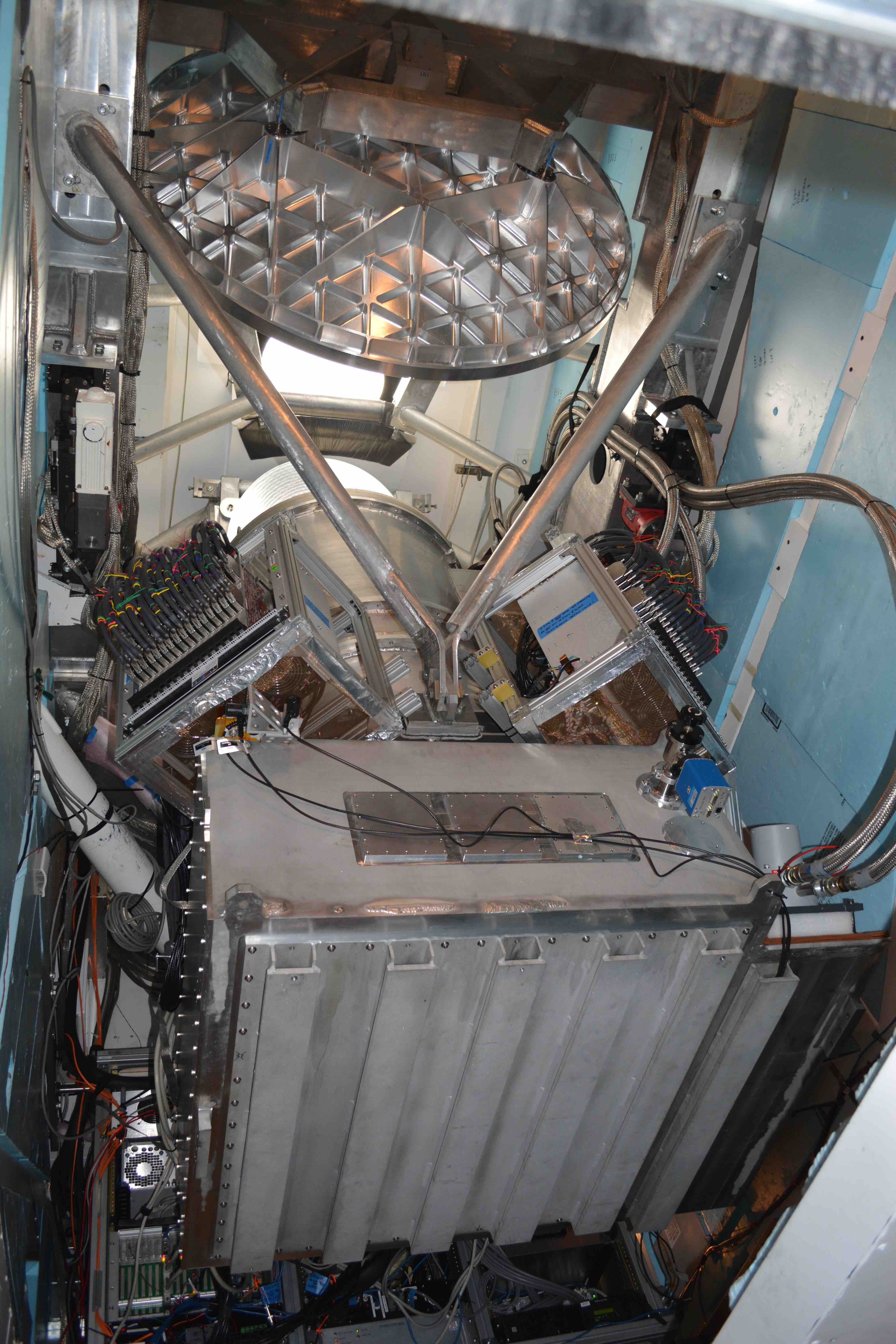}}
\caption{\textit{Left: }The South Pole telescope.  The 10-meter
  primary dish is surrounded by reflective baffles to prevent ground
  contamination.  The SPT-3G receiver is installed in the telescope boom shown
  protruding off the front of the dish.   \textit{Right: } The
SPT-3G cryogenic receiver installed in the telescope boom.  The back side of the
secondary mirror is shown in the upper part of the picture.  Cryogenic
optics are housed in the cylindrical portion of the receiver and the
detectors in the rectangular bottom portion.  Also shown are the two crates
of readout electronics used to operate the detectors. }
\label{fig:spt_pic}
\end{figure}

The South Pole Telescope (SPT) is 10-meter  telescope
surveying the millimeter-wavelength sky from the geographic South
Pole (see Figure \ref{fig:spt_pic})\cite{carlstrom2011}.   
Since its initial commissioning
in 2007, the SPT has hosted three instruments designed to make high
resolution maps of the
cosmic microwave background (CMB).  
The CMB intensity and polarization data from SPT-SZ and SPTpol have been used  for a
number of cosmological and astrophysical studies.  These include  measurements
of the CMB power spectrum damping tail\cite{story2013, henning2018},
reconstructing maps of CMB gravitational lensing for large-scale
structure (LSS) studies\cite{omori2017,story2015, mocanu2018}, and
searching for galaxy clusters\cite{bleem2015,dehaan2016} and millimeter-wavelength transient events\cite{whitehorn2016}.

SPTpol was replaced with the SPT-3G receiver in late 2016.  The SPT-3G receiver
observes at 95, 150, and 220 GHz with polarization
sensitivity and is a significantly more sensitive
instrument than SPTpol.   
All three SPT instruments use superconducting
detectors that are uniquely sensitive due to their ability to operate in the
photon-noise dominated regime.  The SPT-3G focal plane
hosts 16,140 of these detectors, an order of magnitude
more than SPT-SZ or SPTpol.   With this state-of-the-art focal plane, SPT-3G is opening a new regime in high-precision measurements of the CMB polarization.

The SPT-3G receiver is now in its second year of operation.  A summary
of on-sky performance from the initial commissioning of the instrument can be
found elsewhere\cite{anderson2018,pan2018,everett2018}.  During the austral
summer of 2017-2018 several key changes were made to the receiver to
improve total instrumental sensitivity.  These improvements included new broadband
anti-reflection coatings on the optics, optimized detector thermal
properties, and modifications to the readout electronics to reduce noise.  Here, we discuss the current
performance and status of SPT-3G after the successful implementation of those improvements.   The
structure of the paper is as follows.  In Section \ref{sec:science},
we briefly review the cosmological physics that SPT-3G probes using
observations of the CMB.  Section \ref{sec:instrument} describes key
features of the instrument design and the improvements made to the
receiver for the 2018 observing season.  The overall performance of
the SPT-3G receiver is discussed in Section \ref{sec:performance} and
cosmological forecasts are presented in Section \ref{sec:forecasts}.
  
\section{Scientific Goals}
\label{sec:science}

The SPT-3G receiver measures both the intensity and polarization of
the CMB with exceptional sensitivity at a broad range of angular
scales on the sky.  Here, we briefly review the cosmological physics
targeted by SPT-3G\cite{benson2014}. Using the CMB polarization, SPT-3G will probe
fundamental physics in both the early and late universe.  The CMB is
intrinsically polarized from the Thomson scattering of photons
directly prior to recombination.  The resulting polarization pattern
exhibits even-parity (i.e., it is curl-free), and is referred to as the
\textit{E-mode} polarization \cite{hu2002}.  The CMB E modes originate from
the same primary anisotropies seen in the intensity,
therefore, they provide a semi-independent probe of the base cosmological
model, known as \lcdm.
Precise constraints on the \lcdm model
are well-motivated as hints of tension between local probes and the
CMB have surfaced\cite{freedman2017,henning2018}.  While these
discrepancies might be resolved through unaccounted for instrumental or
astrophysical systematics, they could also be pointing towards new
physics beyond $\Lambda$CDM.  New and independent data
sets such as those from SPT-3G  have the potential to resolve or
widen these cracks. 

The polarization of the CMB also contains an odd-parity component,
known as the $\textit{B modes}$\cite{cmbs4science}.   As the CMB traverses through the universe, the light is gravitationally lensed by the
structure that it encounters.  The lensing introduces a curl into the
polarization pattern,  transforming E-mode polarization into B-mode
polarization.  The amplitude of the lensing B-mode signal is proportional to the
amount of LSS in the universe, providing a probe of
physics that influences LSS growth, including massive neutrinos.
Additionally, gravitational waves from an inflationary epoch imprint a
B-mode polarization pattern in the CMB. 
Detection of this signal
would provide
extremely compelling evidence in support of inflation, as well as
constraining its energy scale.  Measurements of primordial
gravitational waves are
parameterized in terms of $r$ (the tensor to scalar ratio),  which
corresponds to the amplitude of the B-mode signal.  Inflationary B modes are undetected at this point in time with the most stringent upper limit placed by BICEP2/Keck of $r <$ 0.07 at 95\% confidence\cite{bicepkeck2016}.

Measurements of the polarization of
the CMB, and particularly the B modes, will provide significant breakthroughs in our fundamental understanding of
both inflation and the neutrino sector.  There are, however, three
sizeable barriers to achieving these landmark measurements.  First,
the gravitational lensing B-mode polarization signal is roughly four to five orders
of magnitude fainter than the CMB intensity anisotropies and the primordial
signal is constrained to be even smaller.  A significant leap in instrumental sensitivity is required
to improve on current measurements.  Second, galactic foregrounds
confuse the CMB signal with both synchrotron and dust
components\cite{bicepkeckplanck2015, planck2018IV}.  Observations
at multiple frequencies are necessary to perform component separation.
Finally, the gravitational lensing B modes occur mostly on small
angular scales and the inflationary B-mode signal is at larger angular
scales.  However, there is a significant region of overlap where the
lensing B modes are essentially a foreground for the inflationary
signal.  Cleaning the lensing B modes from the data (a process called
delensing) will be critically important in improving on current constraints
on inflationary B modes\cite{carron2017,manzotti2017}.  The SPT-3G
instrument is designed with all three of these challenges in mind.
The large-format focal plane has an order of magnitude more
detectors than SPTpol,
significantly increasing instrumental sensitivity.  SPT-3G observes at
three different frequency bands to facilitate component separation between
the CMB and galactic foregrounds. Lastly, SPT-3G maps the CMB with 
high resolution.  These data will result in a high signal-to-noise
measurement of the gravitational lensing B modes and their use in the
delensing process.

In addition to the cosmological signals discussed above, the
high resolution of the SPT enables scientific studies involving
detection of galaxies and galaxy clusters.
Clusters of galaxies are detected through the
scattering of CMB photons off hot (keV) electrons in the intracluster
medium, known as the Sunyaev-Zel'dovich effect (SZE)\cite{sunyaev72}.  Because the SZE
is a scattering effect it does not suffer from the cosmological
dimming that makes x-ray and optical observations of high-redshift
clusters challenging.  The resulting sample of galaxy clusters from SPT-3G
will be nearly mass-limited and extend over a wide range of
redshifts. The clusters serve as a tracer of the evolution of LSS in
the universe and thus a probe of neutrino mass and dark
energy\cite{carlstrom2002}.   SPT-3G will also detect high-redshift
dusty star-forming galaxies, which will be used in studies of 
galaxy formation in the early universe\cite{vieira2010}.  

\section{The SPT-3G Instrument}
\label{sec:instrument}
The SPT-3G instrument is designed to couple the maximum number of detectors to the sky allowed by the existing SPT
primary mirror and receiver cabin.  The instrument consists of an optical chain of both
 mirrors and lenses, a focal plane of detectors,
and the associated readout electronics.  Each component has unique
requirements and challenges, which are briefly summarized in the
following sections.   The SPT-3G instrument was installed into
the telescope in early 2017 and spent the first year of operations focused on
commissioning and calibration \cite{anderson2018, pan2018}.  Based on the results of these data,
three improvements to the receiver in the areas of optics, detectors, and readout
were implemented.   The
motivation and features of these improvements are discussed below. 

\subsection{Optics}

\begin{figure}[t]\centering
\subfigure{\includegraphics[height=0.23\textheight]{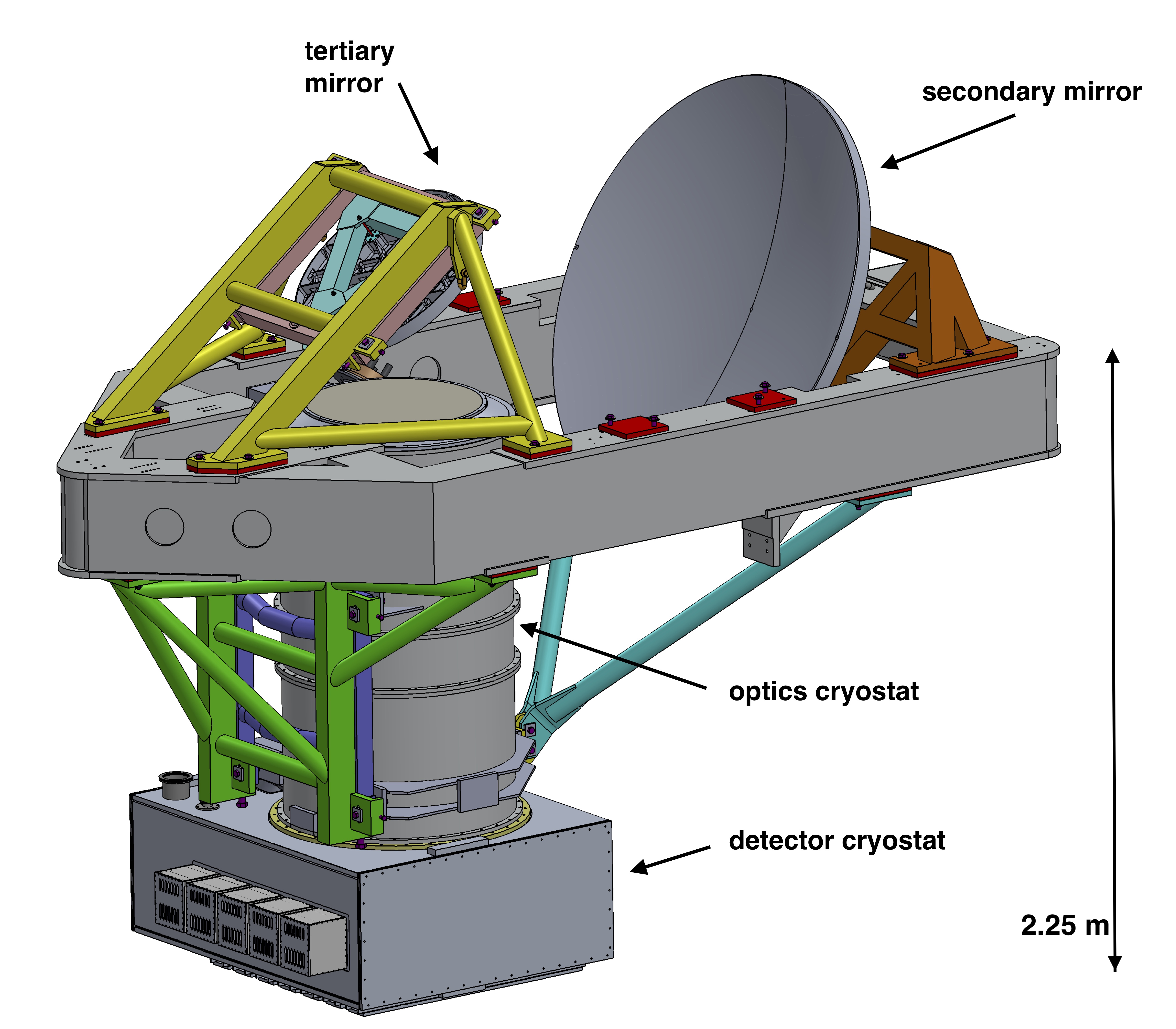}}
\hspace{0.03\textwidth}
\subfigure{\includegraphics[height=0.23\textheight]{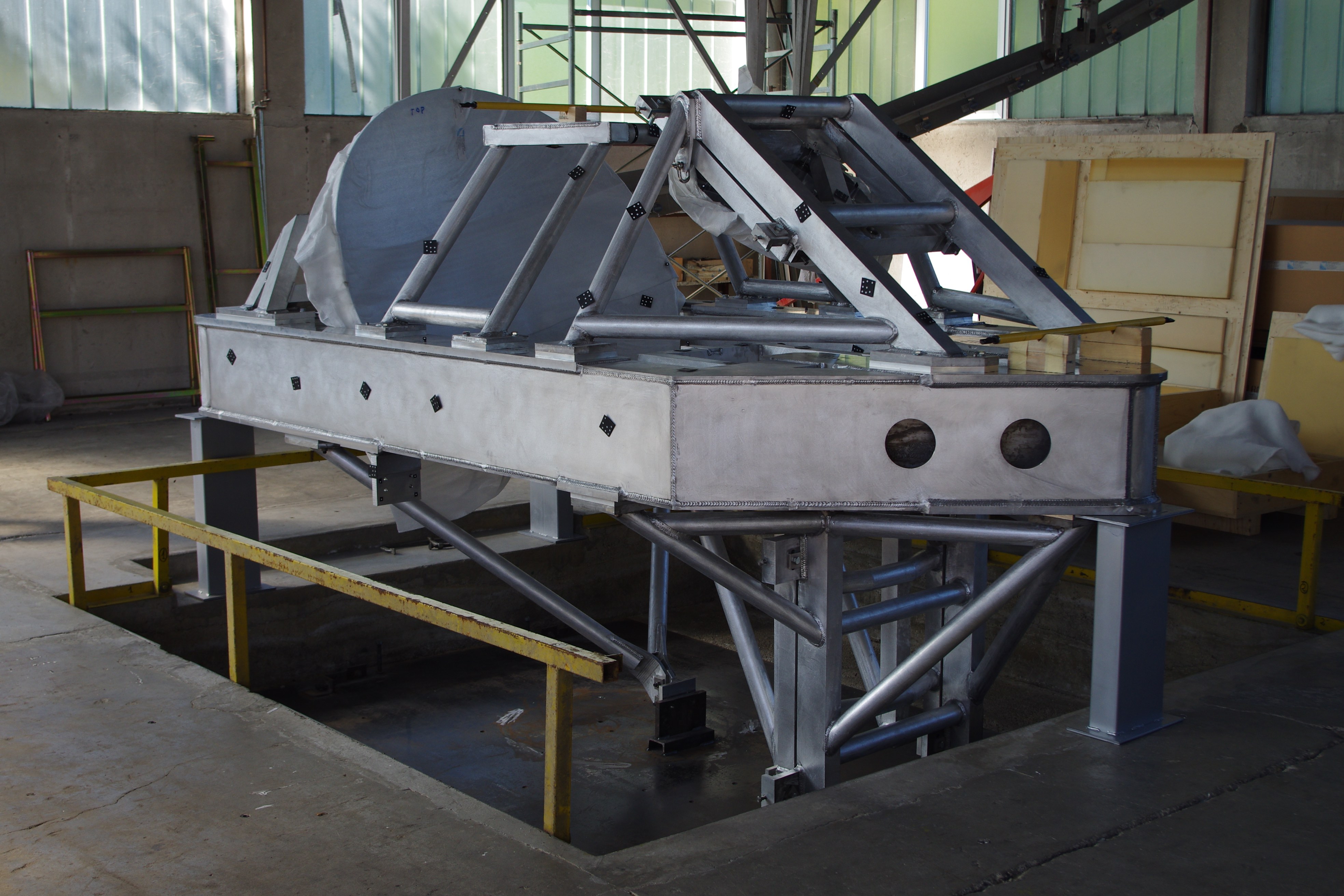}}
\caption{\textit{Left:} A CAD drawing of the SPT-3G
  optics and cryogenic receiver configuration. \textit{Right:} The
  SPT-3G optics bench, including the secondary and tertiary mirrors.}
\label{fig:spt3g_optics}
\end{figure}

The SPT-3G instrument couples light from the 10-meter primary mirror to the
detectors using a set of wide-field broadband optics\cite{benson2014, nadolski2018}.  An ellipsoidal
secondary and flat tertiary mirror (both at ambient temperature)
direct the light from the primary dish through a window into the
optics section of the
receiver cryostat (see Figure \ref{fig:spt3g_optics}).   From there the light passes through an alumina
plate that provides support for the vacuum window and three large (720 mm diameter)
cryogenic alumina lenses.  Additionally, there is a 280 mm
Lyot stop which reduces the illumination of the primary
mirror to the inner eight meters.  The optical configuration results
in a 1.9 degree field-of-view and a large 43 cm diameter image plane.
Hemispherical alumina lenslets couple the microwave light from the image plane
into antennas in the detector array.  With diffraction-limited
performance,  a beam size of 1.6/1.2/1.1 arcminutes is measured at
95/150/220 GHz, respectively. 

The wide-field design of the SPT-3G optics enables the packing of
16,140 detectors into the focal plane and a high array sensitivity
compared to its predecessors.  In
addition to this wide field, a
second key consideration in the optics design is to ensure high
optical transmission  across the 95, 150, and 220 GHz observation bands.
Transmission at the surface of each optical element
directly impacts total instrumental sensitivity.  The vacuum window is made from annealed HDPE, which is both highly
transparent to microwave radiation and strong enough to support the
approximately 12,000 pound load from the atmosphere.  Instead of an
antireflection coating, 1.321 mm tall triangular grooves
spaced at 0.609 mm intervals are
machined into each side of the window.  The grooves are orthogonal on
opposite sides of the window to eliminate any net polarization
rotation.  A previous implementation of a similar groove geometry
resulted in excellent transmission at the relevant
wavelengths\cite{raguin1993} and 95\% transmission was
verified using a fourier transform spectrometer. 

The alumina lenses and window backing plate are anti-reflection (AR) coated to minimize
reflections at these surfaces, increasing total transmission
and minimizing thermal loading from stray light scattering inside the receiver.  For the initial deployment of SPT-3G,
the large lenses had a two-layer coating optimized for the 95/150 GHz
bands that was applied using plasma spray technology\cite{jeong2016}.   Due
to challenges encountered with the performance of the third layer (220
GHz) at cryogenic temperatures the AR coating technology was updated
for the second year of operation.  New lenses with a three-layer
polytetrafluoroethylene-based (PTFE) coating were manufactured and
installed\cite{nadolski2018}. The alumina lenslets (shown in the left panel of Figure
\ref{fig:spt3g_focalplane}) are also AR coated with three layers of
PTFE.   

\subsection{Detectors}

\begin{figure}[t]\centering
\subfigure{\includegraphics[height=0.21\textheight]{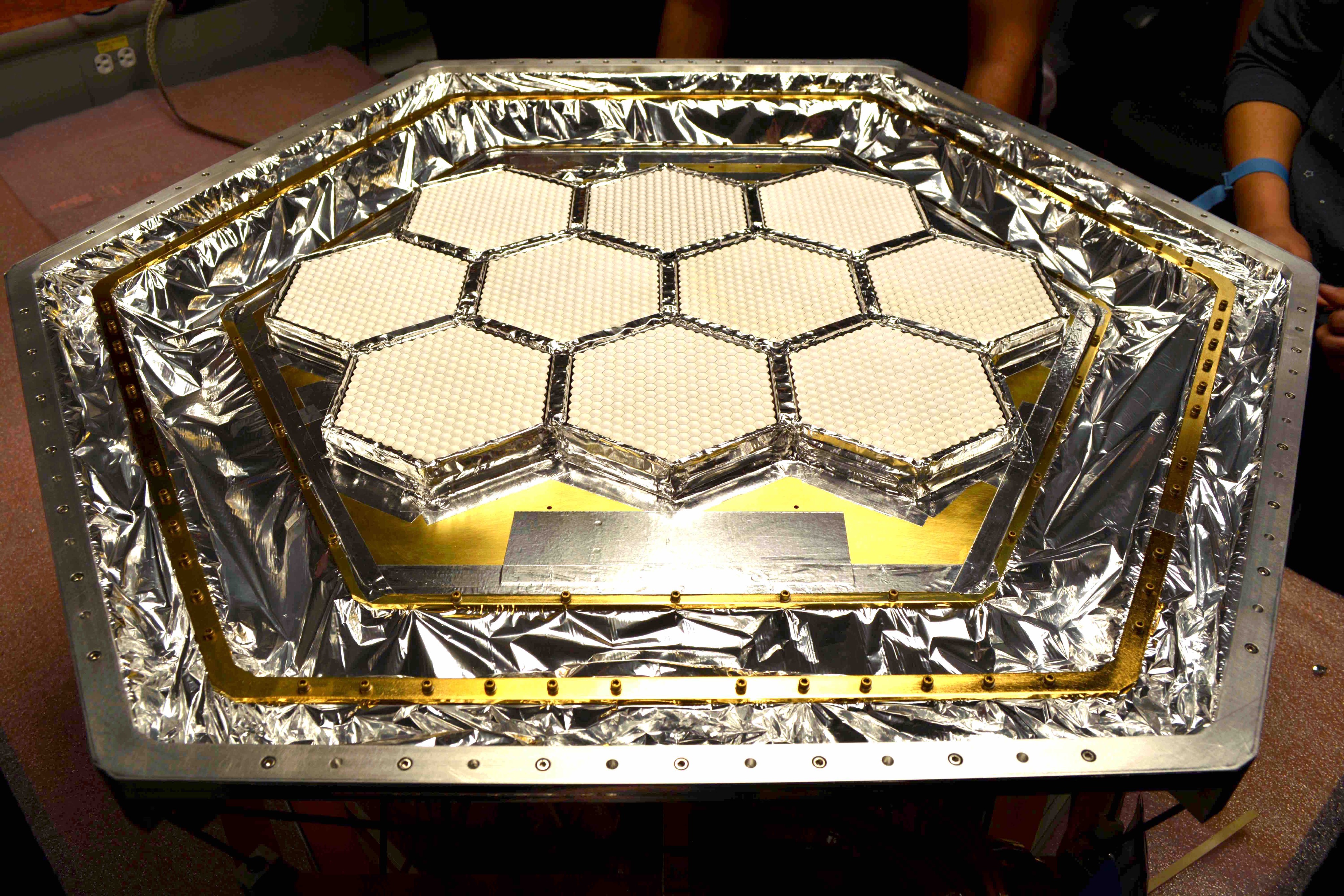}}
\hspace{0.005\textwidth}
\subfigure{\includegraphics[height=0.21\textheight]{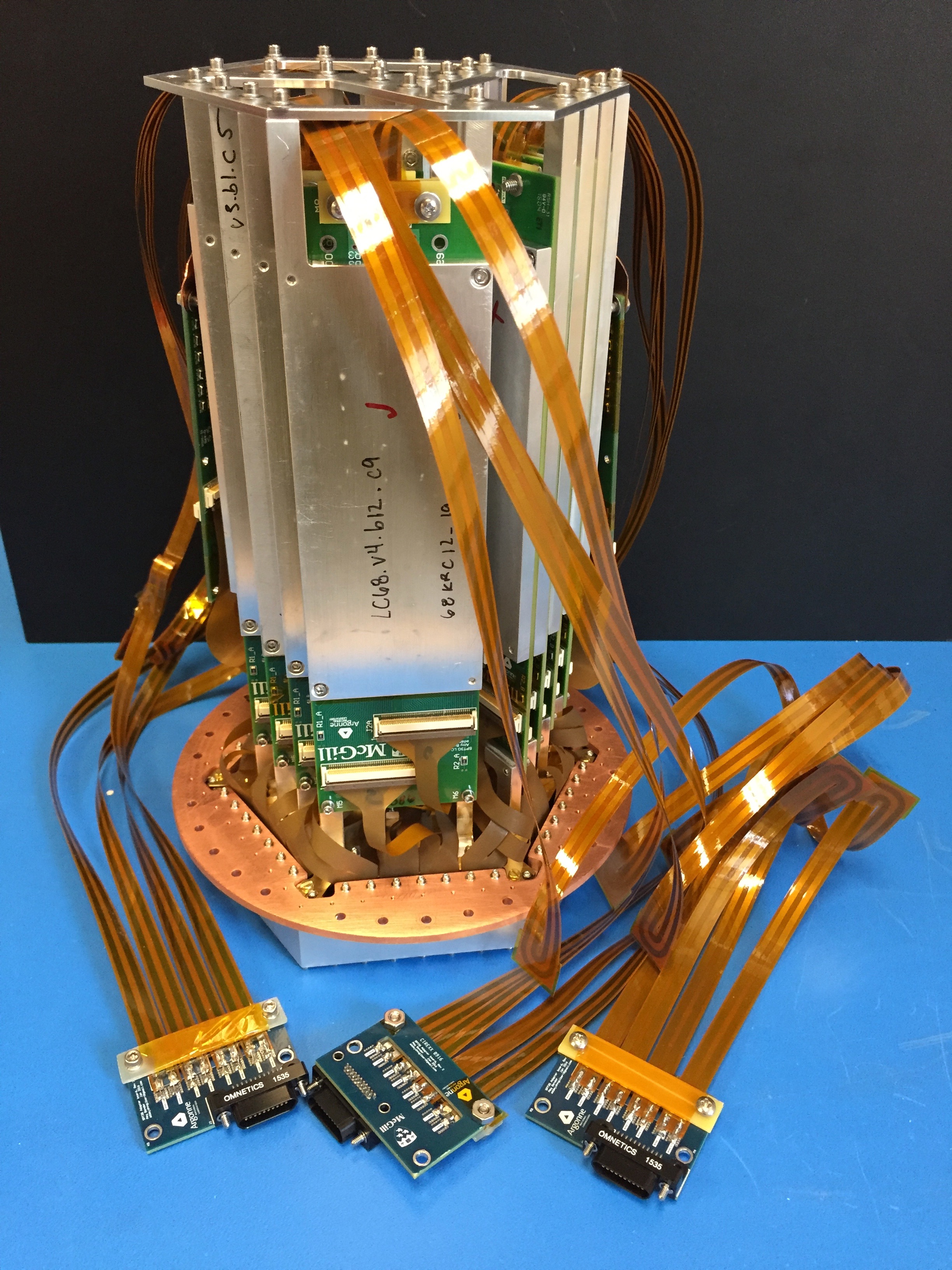}}
\hspace{0.005\textwidth}
\subfigure{\includegraphics[height=0.21\textheight]{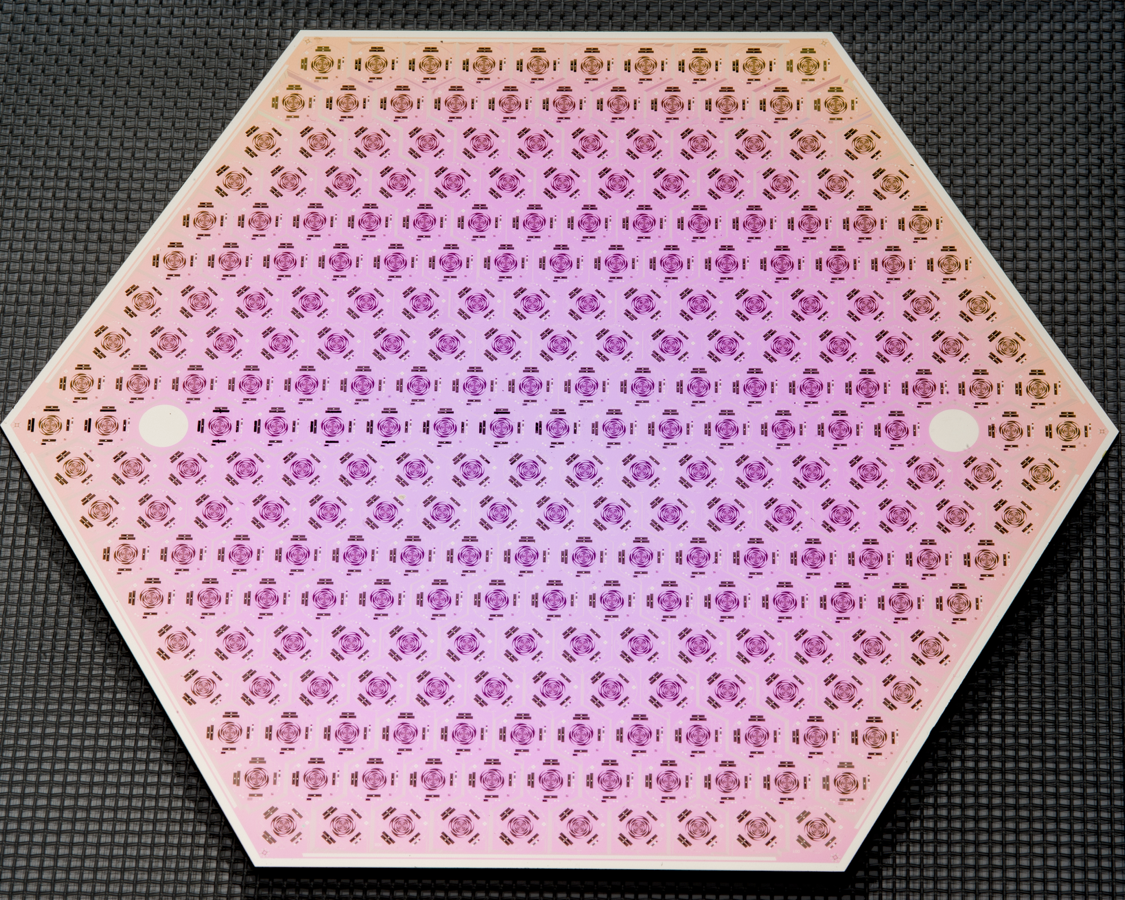}}
\caption{\textit{Left:} The fully assembled SPT-3G focal plane.  In
  the center are the ten hexagonal detector modules.  The white
  hemispherical lenslets couple the microwave power to antennas
  below. Also shown is aluminized mylar that serves as an RF shield
  connecting between the different temperature stages of the focal
  plane structure. \textit{Middle:} The backside of a single detector
  module with the millikelvin readout electronics (the $LC$ filter
  networks)  assembled onto the backplate. The long broadside-coupled striplines
  connect the module to the SQUID amplifiers.  \textit{Right:}  An
  SPT-3G detector wafer with 269 pixels. Each of the $\sim$3 mm
  antennas is connected to and surrounded by six transition-edge
  sensor bolometers.} 
\label{fig:spt3g_focalplane}
\end{figure}

The SPT-3G focal plane contains pixels that are sensitive to both orthogonal
linear polarizations in three observing bands (95/150/220 GHz).
Each pixel contains six individual detectors to measure these signals.
Microwave radiation is absorbed by a broadband sinuous
antenna, measuring both polarizations
  simultaneously in perpendicular arms and then directing the power
  onto separate microstrip lines\cite{obrient2013}.  The power is further separated by observing band using an
  in-line triplexing filter and then deposited onto transition-edge
  sensor (TES) bolometers.  The TES is made of a quad-layer stack of
  titanium-gold-titanium-gold with an overall superconducting transition
  temperature $T_c \sim 420 - 480$ mK\cite{carter2018}.  Additional details on the SPT-3G pixel design and its
  performance can be found in previous publications\cite{posada2016,ding2017,posada2018}.
  The detectors are fabricated on six-inch
  silicon wafers each containing 269 pixels (see the right hand panel
  of Figure \ref{fig:spt3g_focalplane}).  There are ten of the
  hexagonal detector wafers in the SPT-3G focal plane for a total of
  16,140 TES bolometers.  The detectors and a portion of the
  associated readout are cooled to a temperature of $\sim$280 mK
 using closed-cycle mechanical refrigerators. 

The sensitivity with which a TES can map the sky is determined by its
noise equivalent power (NEP).  In the ideal photon-noise dominated limit,
the intrinsic NEP of the TES itself and the NEP of the readout system
are subdominant to the fluctuations in arrival of the incoming
microwave photons.  The level of thermal carrier (phonon) noise in the TES is
related to the saturation power ($P_{\mathrm{sat}}$, the amount of power
required to drive the TES into a normal resistive state).  The \psat for the
SPT-3G detectors has been optimized to reduce this noise contribution
while still maintaining dynamic range and stability requirements.  Beyond the improvement to
the phonon noise, reducing \psat reduces the electrical power and thus
the voltage bias needed to operate the TES stably in the
superconducting transition.  The noise power contributed by the readout
system is proportional to the voltage bias, therefore, optimized \psat
benefits the the readout noise as well (see Section \ref{sec:readout}).

In the first year of operation, the saturation powers in typical SPT-3G
detectors were 17/22/25 pW for 95/150/220 GHz, respectively\cite{anderson2018}.
These values were higher than optimal due to uncertainty in the optical
loading \cite{anderson2018,ding2018}.   Additional detector wafers were fabricated
targeting lower \psat  in all three bands.  Ten new detector wafers (the
full SPT-3G focal plane) were installed into the SPT-3G receiver for
the second year of operation with a median \psat of roughly 12/14/14
pW\cite{dutcher2018}.  

\subsection{Readout}
\label{sec:readout}
\begin{figure}[t]\centering
\subfigure{\includegraphics[height=0.2\textheight]{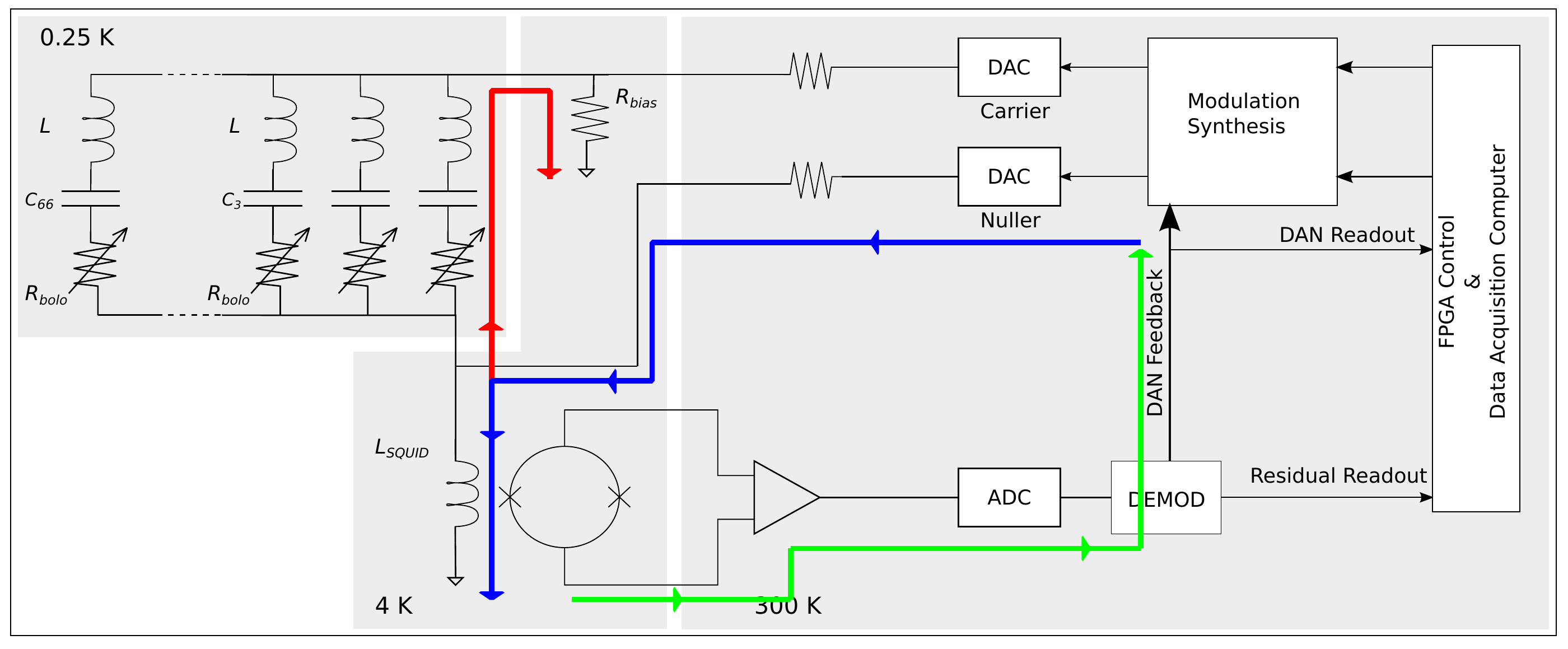}}
\hspace{0.01\textwidth}
\subfigure{\includegraphics[height=0.2\textheight]{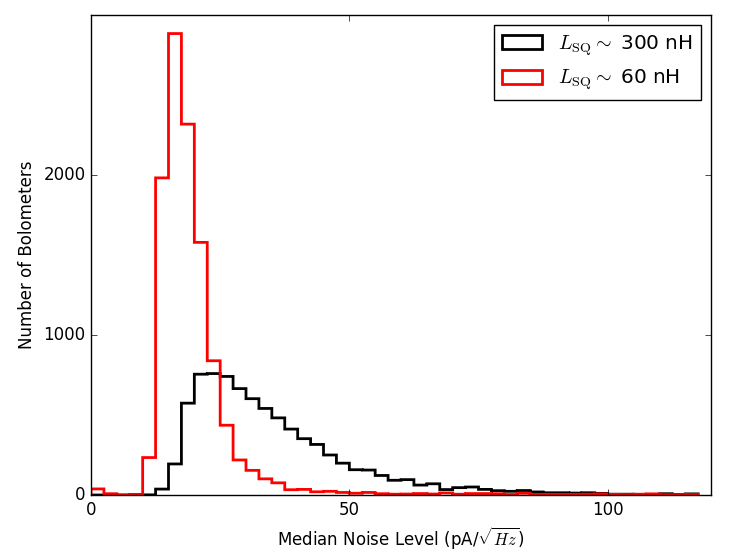}}
\caption{\textit{Left:} A schematic diagram of the fMux readout system
  as implemented in SPT-3G. The blue and red arrows demonstrate the two
  possible paths that current can flow when injected by the nuller,
  creating an amplification of noise that couples into the system in
  the demodulation chain (green arrow, see text).  \textit{Right:}  A comparison
of SPT-3G noise performance when the bolometers are held above the
superconducting transition temperature.   The black curve shows the noise prior to the
improvement in the filtering of signal waveforms output by the
FPGA motherboard and the installation of SQUIDs with lower input
impedance to mitigate demodulation noise amplification. The red histogram
($L_{\mathrm{SQ}} \sim$ 60 nH)  shows the current noise performance of
SPT-3G and the resolution of the excess noise.}
\label{fig:noise_explanation}
\end{figure}

The SPT-3G focal plane is read out using a frequency domain
multiplexing (fMux) scheme\cite{dobbs2012}. Multiplexing is a key technology for
large focal planes like SPT-3G because it reduces the
thermal loading on the millikelvin stage due to readout wiring.  In
fMux, each TES bolometer is connected in series with an inductor and
capacitor to assign it a unique resonant frequency.  These elements are connected in parallel to create a network of resonant filters and detectors (see Figure
\ref{fig:noise_explanation}).    In this configuration a single pair
of wires is used to apply a waveform of AC voltage biases (referred to
as the \textit{carriers}) to the bolometers and
read out the subsequent signals.   The $LC$ filter selects the
appropriate voltage bias for each detector in the network.  In the SPT-3G focal plane, each
module multiplexes 66 bolometers together with resonant frequencies between 1.6
and 5.2 MHz for a total of 240 readout
modules\cite{bender2014, bender2016}.     

When microwave power from the sky is deposited onto the TES island the electrical
resistance and current flowing through the TES changes in response.
The resulting amplitude modulation of the AC voltage bias creates
signals in the sidebands.  The currents from all the TESs in a module
are summed together and input into a Superconducting Quantum
Interference Device (SQUID)\cite{huber2001} where they are amplified, converted to a
voltage, and subsequently digitized.  A process called digital active
nulling (DAN)\cite{dehaan2012} removes both the carriers and the
measured sky sideband signals from the current input into the SQUID (the \textit{nuller} in Figure
\ref{fig:noise_explanation}).  When DAN is running the sideband
signals in the nuller are the final measured sky signal. 

The $LC$ filter networks are constructed through similar
micro-fabrication techniques as the TES detectors.  Interdigitated
capacitors and spiral inductors are patterned in a superconducting
material (either aluminum or niobium) on a silicon
wafer\cite{rotermund2016}.  A fully assembled SPT-3G wafer module is
shown in Figure \ref{fig:spt3g_focalplane}.  Each filter network
is contained in a single chip and connected to the TES detector
wafer using flexible kapton over copper cables and a printed circuit
board.  The filter chips are connected to the SQUID and bias
electronics using superconducting broadside-coupled striplines
\cite{avva2018}.   
  
There are two key parameters of the readout system that directly impact total
instrument sensitivity: readout yield and readout noise.  Crosstalk
is also an important systematic effect when evaluating the performance of the fMux readout
system.  A discussion of crosstalk and the relevant design and
features implemented in the SPT-3G readout are found elsewhere
\cite{bender2016, avva2018}. Yield will be discussed in Section \ref{sec:yield} in the context of
the total focal plane yield.   Here, we focus on the contribution of
readout noise to the total noise budget.   The noise of the \fmux readout
system is measured in terms of the noise equivalent current
($NEI_{\mathrm{readout}}$) at the bolometer.  Transforming this noise
level to the readout noise equivalent power requires scaling by the voltage bias
($NEP_{\mathrm{readout}} = V_{\mathrm{bias}}NEI_{\mathrm{readout}}$).
Noise is expected from several sources in the fMux system including
the signal generation and processing electronics, the resistor used to create the
voltage bias, and the intrinsic SQUID noise\cite{montgomerythesis,dobbs2012}.   

During the first year of operation an excess of readout noise was
observed in SPT-3G, degrading the instrumental sensitivity compared to
expectations.  Three improvements were made to the SPT-3G system to
resolve this excess.  First, a non-optimal instrument grounding scheme
was discovered.  Noise was entering the highly-sensitive analog ground
plane of the SQUID control boards via the shielding braid and foils in the cables that connect to the ADC/DAC mezzanines
and field programmable gate array (FPGA) motherboards.   These ground connections
were disconnected and the readout and instrument noise was
dramatically reduced across the entire bandwidth of the electronics.  Second, filters on the waveforms output by the
carrier and nuller digital-to-analog converters were tightened to
better match the actual bandwidth used.    This reduced the bandwidth
over which external signals could couple to the SQUIDs.

The final source of excess noise in the SPT-3G readout system was a subtle feature of the nulling scheme.   DAN injects and
actively adjusts a waveform at the input coil of the SQUID  to
minimize the voltage at this node.   However, the current is also able
to flow back through the TES/ filter network and bias resistor to
ground (see Figure \ref{fig:noise_explanation}).   Noise sources
introduced into the fMux system after the input coil
of the SQUID are measured in the demodulation process and
added to the injected nuller.  
The nuller increases in amplitude to compensate for the current
leakage, amplifying the demodulation chain noise in the process. 
The amount of current that flows through the secondary path
is determined by the ratio of impedances between the TES and the SQUID input
coil.  The effect is therefore proportional to the TES
bias frequency, with a negligible contribution at the lowest frequencies but rising
to a significant level in the middle of the SPT-3G electronics bandwidth. The typical operating  TES resistance in SPT-3G is
roughly $R_{\mathrm{TES}} \sim 1.6\, \Omega$.  In the first year of
operation the SPT-3G fMux used NIST fabricated SQUIDs (SA4 style) with a typical input coil
inductance of $L_{\mathrm{SQ}} \sim 300$ nH \cite{huber2001}.  For the 1.6 - 5.2 MHz
bandwidth the SQUID input coil impedance becomes comparable to
$R_{\mathrm{TES}}$, resulting in significant amplification of the
demodulation chain noise.  For the second year of operation, new NIST
SA13 SQUIDs with a reduced input impedance were installed in the SPT-3G receiver.   Additionally, the
printed circuit board that the SQUIDs are mounted on was redesigned to
remove its contribution to the circuit impedance between the nuller
injection point and the SQUID input coil.  With a total input
coil inductance closer to $L_{\mathrm{SQ}} \sim 60 - 80$ nH, less of
the nuller current leaks through the TES network and the noise
amplification is significantly reduced.  

Figure \ref{fig:noise_explanation} compares the  total noise of the
SPT-3G detectors currently to the performance before the implementation of the motherboard
filtering and low input inductance SQUIDs.  For
the data in this plot, the detectors were held at a temperature above
the superconducting transition.  Expected noise sources in this
configuration include Johnson noise from the bolometer and the readout
noise.  The reduction in both the median and scatter in the current noise
distribution demonstrates the successful implementation
of the improvements discussed here and the mitigation of the
excess readout noise.  

\section{On-sky Performance}
\label{sec:performance}
SPT-3G observes several calibration sources as part of its regular
cadence for  array
characterization.  These sources include the
atmosphere, a chopped blackbody behind the secondary mirror, the
galactic HII regions RCW38 and MAT5a, the active galactic nucleus
Centaurus A,  planets (when available from the South Pole), and
quasars.  In this section, we present the performance of the SPT-3G receiver in its
improved second year configuration based on observations of these
calibration sources.

\subsection{Array Yield}

\label{sec:yield}
\begin{figure}[t]\centering
\subfigure{\includegraphics[width=0.325\textwidth]{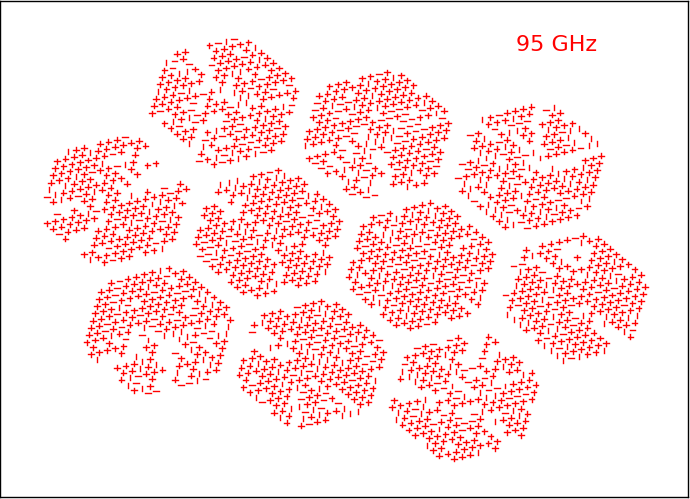}}
\subfigure{\includegraphics[width=0.325\textwidth]{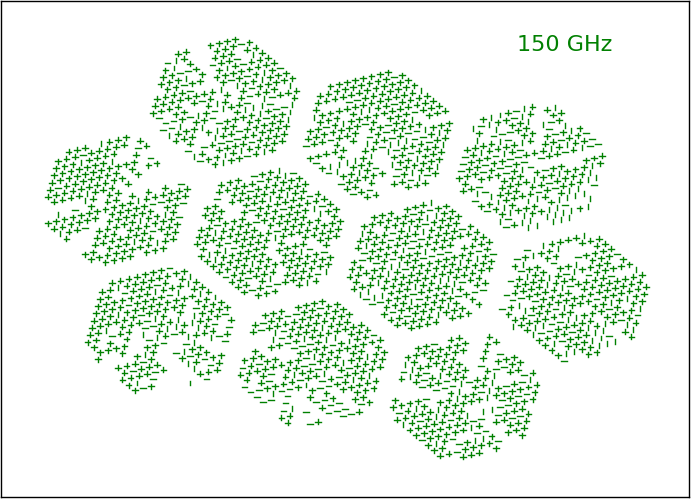}}
\subfigure{\includegraphics[width=0.325\textwidth]{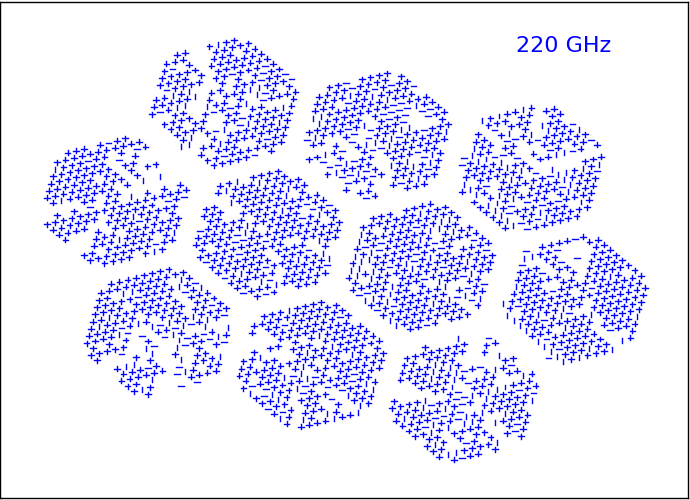}}
\caption{The optically responsive detectors in SPT-3G  as a
  function of their position in the focal plane.   Each detector is
  represented by a line and full polarization pair is shown as a
  $+$.   The hexagonal shapes of the ten detector wafers are clearly visible.  Wedges that
  are missing from a hexagon represent readout modules in the focal
  plane that are electrically disconnected.  These modules will be repaired
  during the austral summer of 2018-2019.}
\label{fig:live_detectors}
\end{figure}

The leap in sensitivity of the SPT-3G receiver compared to previous CMB
experiments is made possible by the large number of TES bolometers.  The
total number of detectors sensitive to optical signals is therefore
critical to achieving the targeted science.   Detectors in the array
that respond to optical calibration signals are shown in
Figure \ref{fig:live_detectors}.  Approximately 11,400 bolometers have optical
response and the corresponding losses can be attributed to several
different effects.  A small fraction of the array is not electrically
connected due to an asymmetry in the wiring design of the detector wafer and cryogenic
readout.  Some  detectors fail initial quality control while others do
not perform cryogenically.  Some yield loss can also be attributed to entire modules that have been disconnected in the cryogenic readout
(either a problem with the filter network chip or the stripline
connection).  These modules are seen in Figure \ref{fig:live_detectors} as
wedge shaped holes in the hexagonal wafer pattern.   Overall, the
total yield for typical observations is $\sim 72$\%\cite{dutcher2018}.

\subsection{Array Noise}
\label{sec:arraynoise}

\begin{figure}[t]\centering
\includegraphics[width=0.45\textwidth]{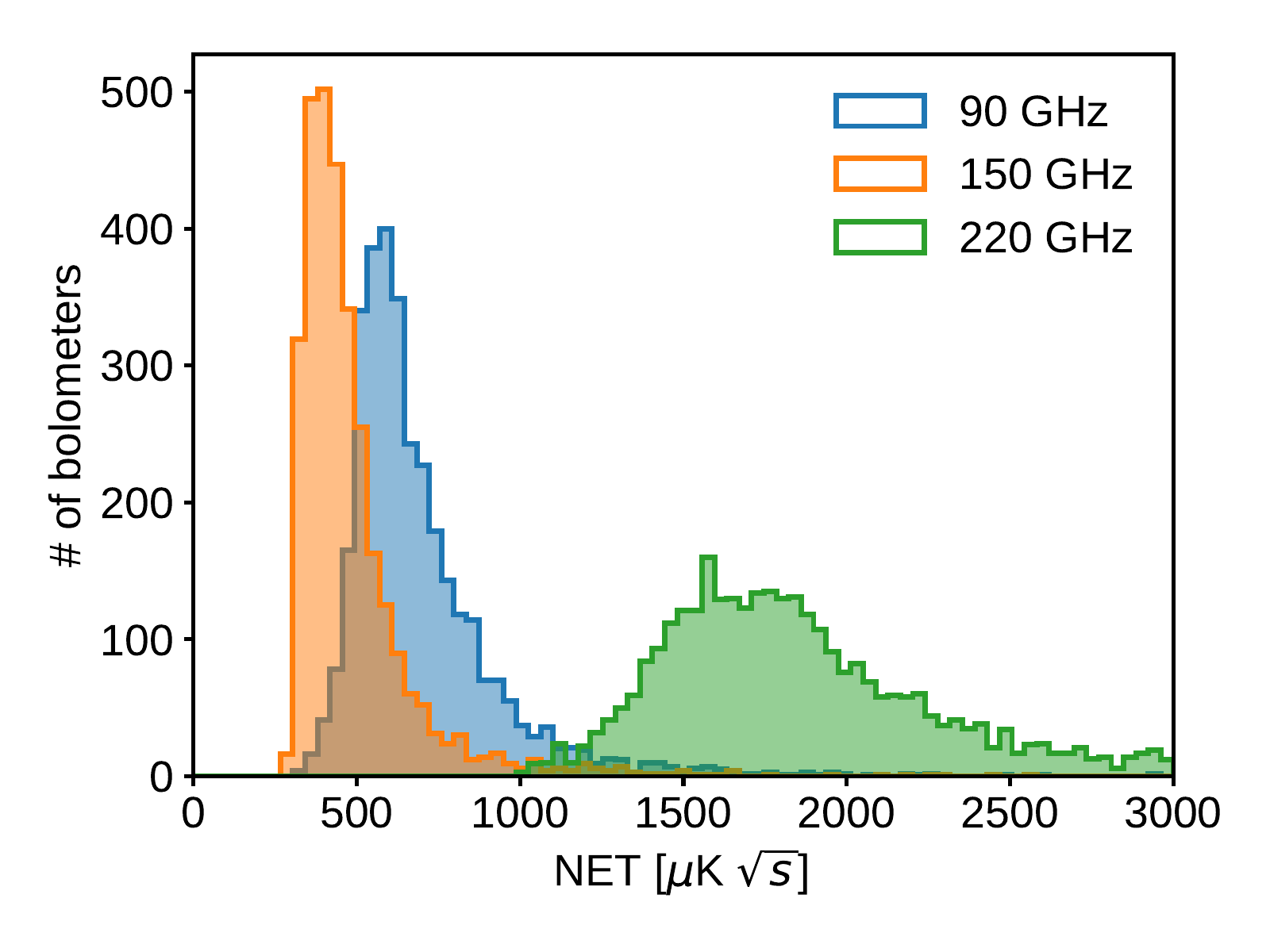}
\caption{The on-sky distribution of noise equivalent temperatures
  (NETs) for the SPT-3G receiver.  The performance of the 95 and 150 GHz detectors
  is in reasonable agreement with expectations, however, performance
  of the 220 GHz detectors is clearly degraded.  This effect is thought
to be due to poor transmission in the AR coatings on the lenses and is
under further investigation.}
\label{fig:net_hist}
\end{figure}

Noise equivalent temperature (NET) is used to quantify the
sensitivity of the detectors including the noise, bands, and optical
efficiency.  NET is determined by measuring the noise in the
10 - 15 Hz band during a noise observation taken while the telescope
is stationary and calibrating it based
on an RCW38 observation.  Current analysis suggests that the white noise
level in this band is representative of the instrument performance at
frequencies of interest.  Analysis of the noise at the lowest
frequencies (relevant for mapping the largest angular scales on the sky) is ongoing.  The distribution of NET for each of the
SPT-3G observing bands is shown in Figure \ref{fig:net_hist} and
summarized in Table \ref{tab:sensitivity}.  We find
median per detector NETs of 630/440/1800 $\mu$K$\sqrt{s}$ in the
95/150/220 GHz bands.    Given the yield stated above, the 
total focal plane NET is 10 $\mu$K$\sqrt{s}$ at 95 GHz,  8
$\mu$K$\sqrt{s}$ at 150 GHz, and 30 $\mu$K$\sqrt{s}$ at 220 GHz.  
The 95 and 150 GHz NETs are reasonably consistent with theoretical
predictions for the SPT-3G optical design, however, the 220 GHz NET is
about a factor of 1.5 higher
than expectations.   
The non-optimal performance of the 220 GHz detectors is
believed to be due to poor transmission in the alumina AR coatings and an
investigation is underway to confirm this hypothesis\cite{sobrin2018}. 
Despite the performance of the 220 GHz bolometers, the SPT-3G array is
significantly more sensitive than SPTpol, with a 9/4
times faster mapping speed at 95/150
GHz\cite{george2012,benson2014}.

{\def\arraystretch{1.5}\tabcolsep=10pt
\begin{table}[htb]
\vskip 12 pt
\small
\begin{center}
\begin{tabular}{l | c | c| c }
\hline
& 95 GHz & 150 GHz & 220 GHz \\
\hline
Beam FWHM (arcmin) & 1.6  & 1.2 & 1.1 \\
Optically responsive $N_{\mathrm{bolo}}$  & 3800& 3780  & 3820   \\
NET$_{\mathrm{bolo}, T}$ [$\mu$K$\sqrt{s}$ ]  & 630  & 440  &  1800 \\
NET$_{\mathrm{array}, T}$  [$\mu$K$\sqrt{s}$ ]   & 10  &  8  & 30  \\
Projected Map Depth [$\mu$K-arcmin]  & 3  & 2 & 9
\end{tabular}
\caption{The measured performance of the SPT-3G receiver as well as projected map depths for the five year 1500
  deg$^2$ survey.}
\label{tab:sensitivity}
\end{center}
\end{table}
}

\subsection{Survey Strategy}
\label{sec:survey}

\begin{figure}[t]\centering
\includegraphics[height=0.36\textheight]{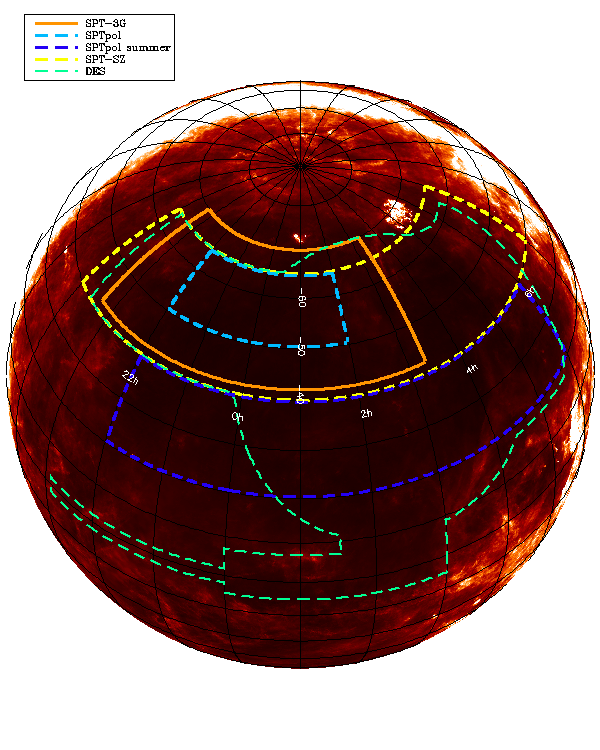}
\hspace{0.03\textwidth}
\includegraphics[height=0.36\textheight]{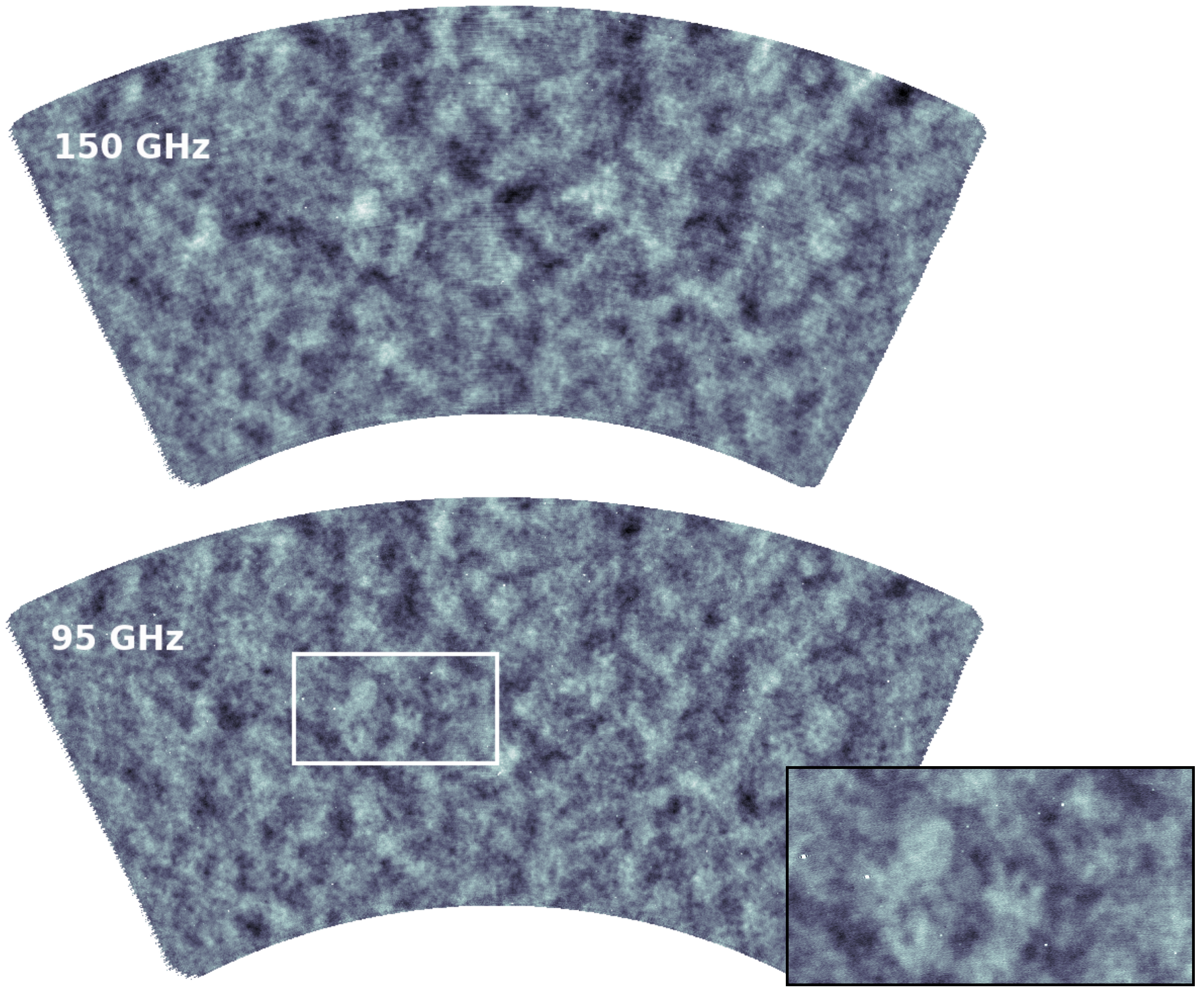}
\caption{\textit{Left:}  The SPT survey fields overlaid on an IRAS
  dust map of the sky\cite{schlegel1998}. In this diagram, the south
  celestial pole is shown pointing up.  The main 1500 deg$^2$ survey field for
  SPT-3G is shown in orange.  The SPTpol 500 deg$^2$ field was 
  observed during SPT-3G commissioning.  The SPT-3G main field is designed for complete coverage
  of the future BICEP Array field.   It also overlaps with existing
  SPT survey data, with the current BICEP/KECK
  field (coincident with the SPTpol
  field)\cite{bicep22014II}, and the DES\cite{desdatarelease2018}.  \textit{Right:}  Intensity maps from SPT-3G commissioning
  observations of the 500 deg$^2$ field. Common structure is clearly
  seen in the larger-scale anisotropies between the two observing
  bands.  Point sources are also visible throughout the map, shown in
  the 95 GHz inset in the lower right image of the panel.}
\label{fig:spt3g_maps}
\end{figure}

 During the
first year of SPT-3G commissioning, observations were taken of the SPTpol 500
square degree field.  An initial intensity map of these data  in the 95 and
150 GHz bands is shown in Figure \ref{fig:spt3g_maps}.   
For its primary science survey, SPT-3G is observing a 1500 square
degree field for a total of five years (from 2018 - 2023).  The survey field, shown in Figure \ref{fig:spt3g_maps},
overlaps completely with the current BICEP/KECK field as well the
future BICEP Array field for joint analyses of
the complementary datasets\cite{bicep22014II}.   The
SPT-3G survey field also has significant overlap with the Dark Energy Survey
(DES) for cross-correlation with optical
tracers of LSS\cite{desdatarelease2018}.  Observations of the 1500 deg$^2$
field are now ongoing and initial analysis of the data is underway.
Given the array sensitivities stated above, final map depths of 3, 2, and 9 $\mu$K-arcmin in intensity are expected in the 95, 150, and
220 GHz bands with polarization depths a factor $\sqrt{2}$ higher. 

The 1500 deg$^2$ field is split into two subfields at high and low
telescope elevation to prevent dramatic changes in the atmospheric optical loading on the
detectors.  Observations of each subfield alternate with recycling the
millikelvin refrigerator (approximately every 26 hours) and
observations of  the internal calibrator and galactic HII regions
(RCW38 and MAT5a) are interspersed throughout.

\begin{figure}[t]\centering
\includegraphics[width=0.95\textwidth]{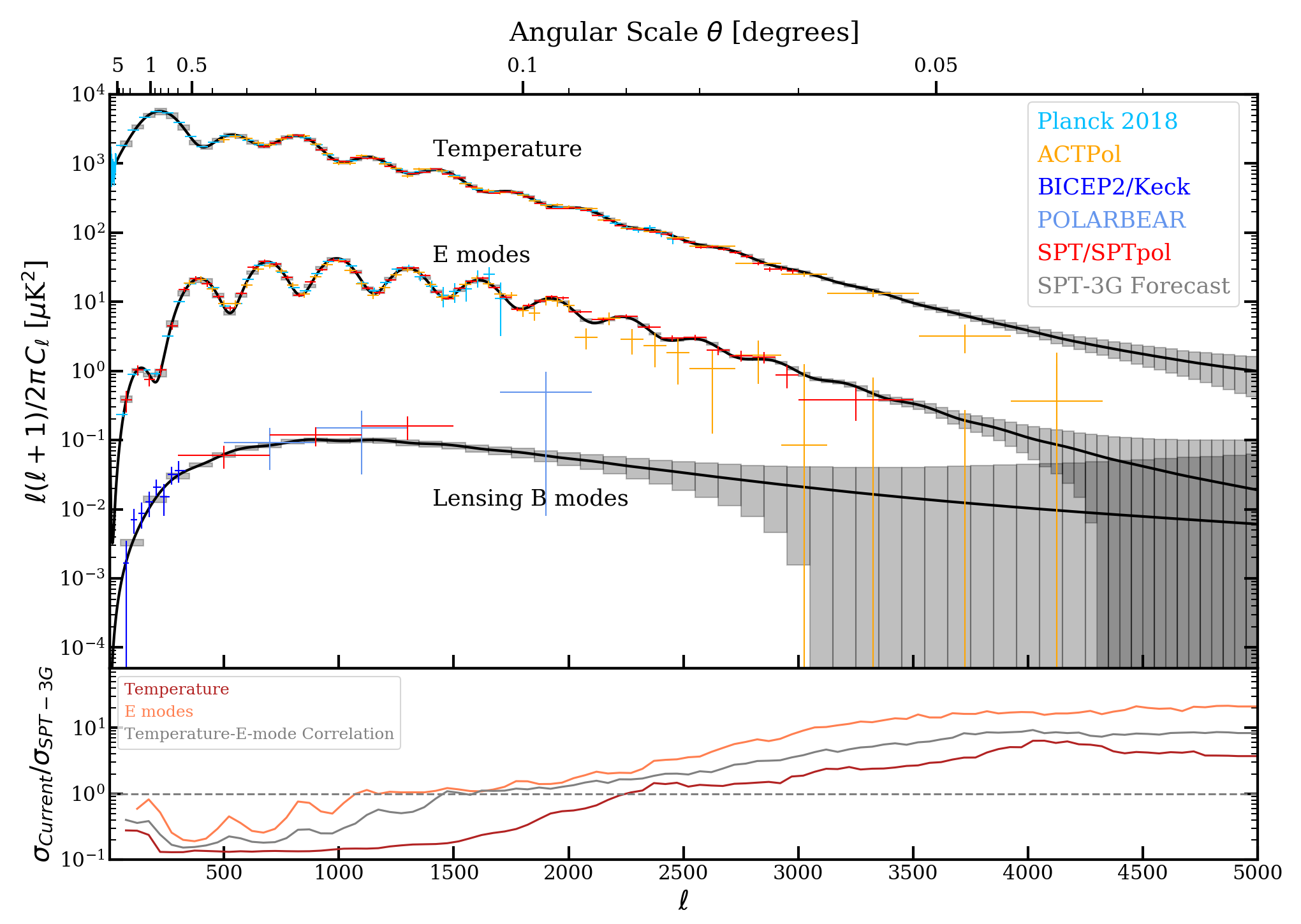}
\caption{\textit{Top:} Projections for the CMB temperature and polarized power spectra
as measured by SPT-3G are depicted by the gray vertical bars.  Current state-of-the-art
constraints from the \textit{Planck} satellite\cite{planck2018VI},
BICEP2/KECK\cite{bicepkeck2016}, ACTPol\cite{louis2017}, POLARBEAR\cite{akiba2017}
SPT-SZ\cite{story2013}, and SPTpol\cite{kiesler2015 ,henning2018} experiments are also shown. \textit{Bottom:} Expected improvement in
the TT, EE, and TE power spectra for SPT-3G data in comparison to the
most sensitive current measurements.  SPT-3G will improve on current
constraints at the multipoles where these factors are greater
than unity (the dashed line).  }
\label{fig:spt_powerspectrum}
\end{figure}

\section{Cosmological Forecasts}
\label{sec:forecasts}
Given the array sensitivity presented in the previous section
projections can be made for the intensity (temperature) and polarized CMB
power spectra as measured by SPT-3G.  Figure \ref{fig:spt_powerspectrum}
shows these projections in comparison with current state-of-the-art
datasets.  These forecasts highlight how SPT-3G will
improve on current constraints, particularly at small angular scales
(high multipole moments) on the sky.  The damping tail measurements of
the E modes will probe both the basic \lcdm model and the effect of light
relics\cite{cmbs4science}.  SPT-3G is expected to constrain the number of light relics with $\sigma(N_{\mathrm{eff}}) =
0.1$ through the impact on the expansion rate of the
universe at early times seen in the CMB power spectrum damping tail.

\begin{figure}[h]\centering
\includegraphics[height=0.27\textheight]{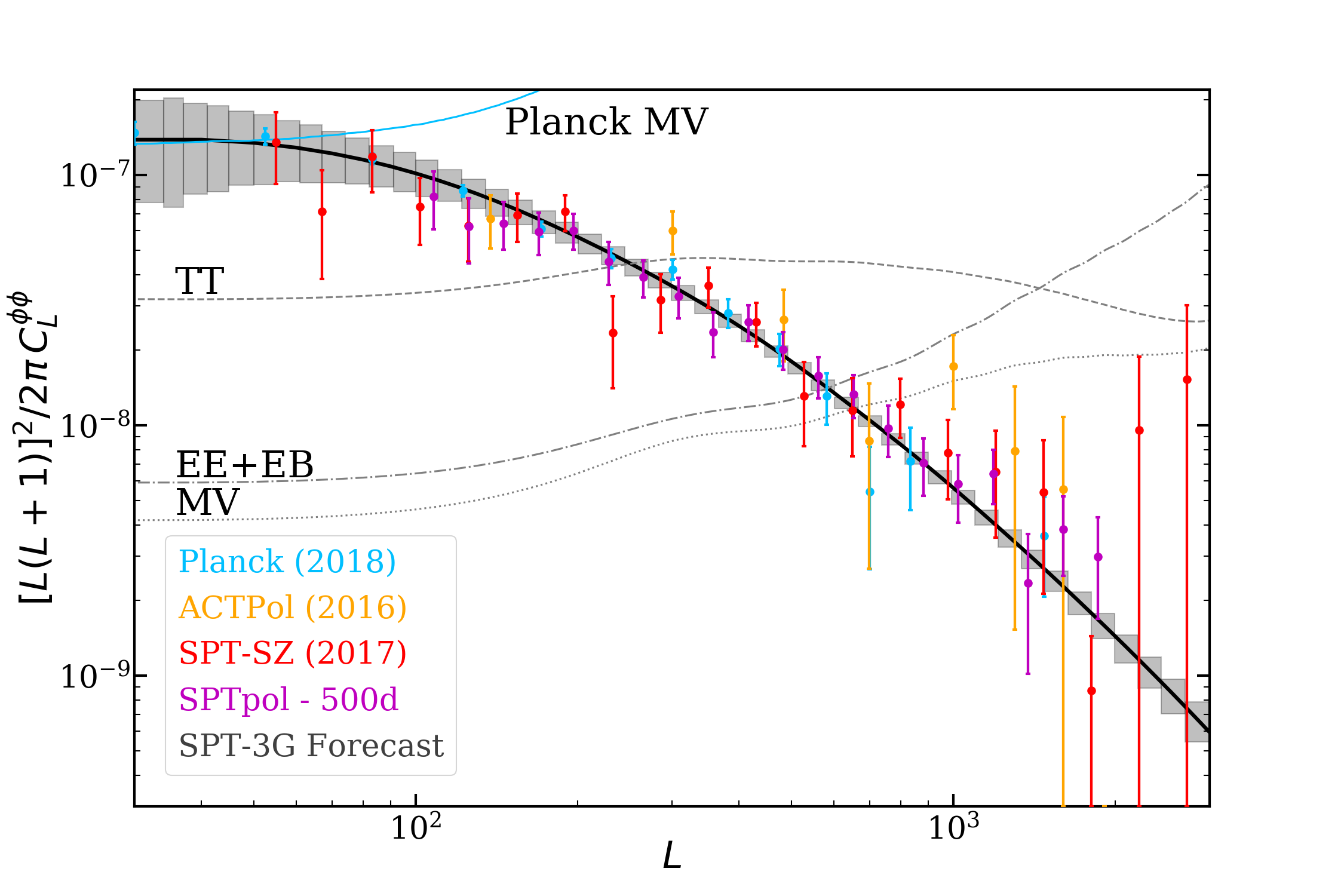}
\includegraphics[height=0.27\textheight]{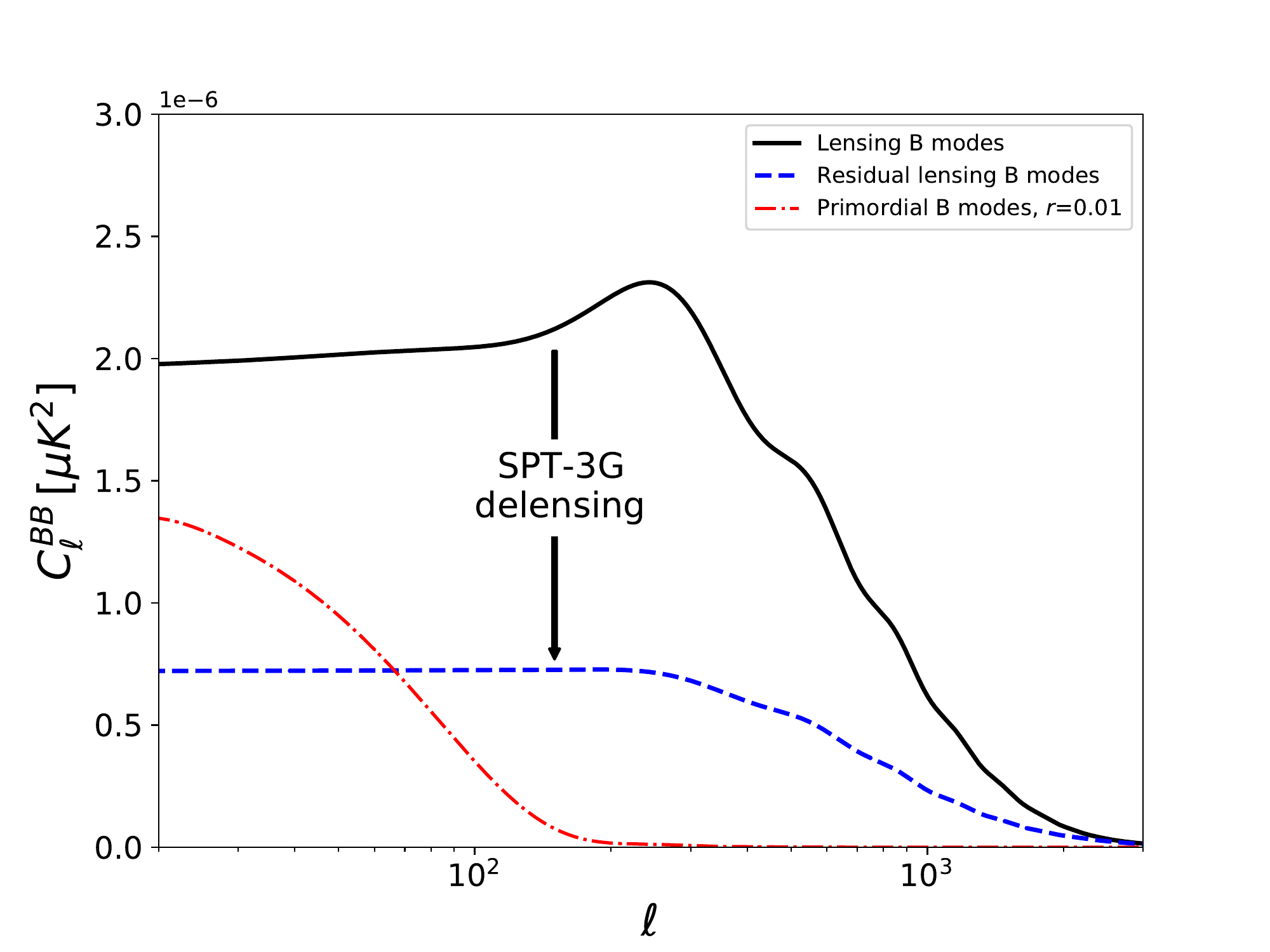}
\caption{\textit{Left:}  Projected SPT-3G measurement of the CMB
  gravitational lensing power spectrum, along with current
  measurements from the \textit{Planck} satellite\cite{planck2018VIII}, SPT-SZ\cite{omori2017}, SPTpol\cite{story2015,mocanu2018}, and
  ACTPol\cite{sherwin2017}. Noise curves are shown for different
  combinations of the lensing estimator (TT only, EE + EB, and the
  minimum variance combination of all temperature and polarization
  estimators).  For lensing modes smaller than the mode at which 
  the two curves cross, SPT-3G will have signal-to-noise greater than
  unity on individual sky features. \textit{Right:} The projected cleaning 
  of the gravitational lensing signal from the B-mode power spectrum
  using the SPT-3G data (black and blue dashed). For comparison, a primordial
  B-mode spectrum due to inflationary gravitational waves is shown.}
\label{fig:spt_lensing}

\end{figure}

The high signal-to-noise measurement of small scale B modes will tightly
constrain the gravitational lensing component of the power spectrum.
The projection for the SPT-3G measurement of the gravitational lensing
power spectrum is shown in Figure \ref{fig:spt_lensing}.   From these
data, SPT-3G will infer the sum of neutrino mass through its impact on
the growth of large scale structure.  The expected constraint on the
sum of neutrino mass is $\sigma (\Sigma m_{\nu}) \sim 0.1$ eV, which is
approaching an interesting regime for differentiating between
possibilities for the neutrino
hierarchy\cite{cmbs4science}.  Additionally, the SPT-3G
measurement of the lensing B modes will be used to delens large-scale
B-mode data in search of inflationary gravitational waves (see Figure
\ref{fig:spt_lensing}).  As described in Section \ref{sec:survey} the
SPT-3G field fully overlaps with that of the BICEP/KECK survey.  SPT-3G
will therefore be able to both internally delens its own measurement
of the large-scale B modes as well as the data from
BICEP/KECK.   It is expected that delensing with the SPT-3G data will
remove approximately 2/3 of the lensing power, tightening constraints on
inflationary B modes.

\section{Summary}

The SPT-3G receiver is a high throughput, broadband (95/150/220 GHz),
polarization-sensitive  TES bolometer camera
designed for high resolution observations of the CMB.  We presented here the
status and performance of SPT-3G in its second year of operations.  Based on the initial year of commissioning
and calibration observations in 2017, several improvements were made to
the receiver for the 2018 observing season.  A new PTFE based
anti-reflection coating technology  was used to apply three-layer
coatings to the large cryogenic alumina lenses.  New detectors with
optimized saturation power were fabricated and installed.  Lastly,
excess noise in the readout system was successfully mitigated by
improving grounding and installing lower input impedance SQUIDs.
With these improvements all three bands show performance
  consistent with photon-noise domination. The 220 GHz band shows some degradation of sensitivity
and an investigation of the optical throughput is underway.   There
are approximately 11,400 optically responsive TES bolometers in the
full focal plane, making the SPT-3G camera one of the most sensitive
CMB cameras currently observing.   SPT-3G has now begun a five year
survey of its primary 1500 deg$^2$ field.  The exceptional sensitivity
of the resulting CMB E- and B-mode polarization maps will lead to a
wide variety of millimeter-wavelength point source astrophysics, tight constraints on neutrino mass, and delensing of the B-mode signal
for inflationary studies. 

\section{Acknowledgements}
The South Pole Telescope program is supported by the National Science Foundation (NSF)  through grant PLR-1248097.
Partial support is also provided by the NSF Physics Frontier Center grant PHY-1125897 to the Kavli Institute of Cosmological Physics at the University of Chicago, the Kavli Foundation, and the Gordon and Betty Moore Foundation through grant GBMF\#947 to the University of Chicago.  Work at Argonne National Lab is supported by UChicago Argonne LLC, Operator of Argonne National Laboratory (Argonne). Argonne, a U.S. Department of Energy Office of Science Laboratory, is operated under contract no. DE-AC02-06CH11357. We acknowledge R. Divan, L. Stan, C.S. Miller, and V. Kutepova for supporting our work in the Argonne Center for Nanoscale Materials.
Work at Fermi National Accelerator Laboratory, a DOE-OS, HEP User Facility managed by the Fermi Research Alliance, LLC, was supported under Contract No. DE-AC02-07CH11359.  NWH acknowledges support from NSF CAREER grant AST-0956135. The McGill authors acknowledge funding from the Natural Sciences and Engineering Research Council of Canada, Canadian Institute for Advanced Research, and the Fonds de recherche du Québec Nature et technologies. CR acknowledges support from the Australian Research Council’s Future Fellowships scheme (FT150100074).  JV acknowledges support from the Sloan Foundation.

\bibliography{bender_spie_2018}   
\bibliographystyle{spiebib}

\end{document}